\documentclass[preprintnumbers, 
floatfix,twocolumn,preprintnumbers, letterpaper, superscriptaddress,nofootinbib]{revtex4}
\usepackage{amsfonts}
\usepackage{amsmath}
\usepackage{amssymb,epsf}
\usepackage{latexsym}
\usepackage{graphicx,epsfig}
\usepackage{amssymb}
\usepackage{float}
\usepackage{subfigure}
\usepackage{epstopdf}
\usepackage[colorlinks=true,citecolor=blue,linkcolor=blue,urlcolor=black]{hyperref}
\usepackage{dcolumn}
\usepackage{psfrag}
\usepackage{wrapfig}
\usepackage{makeidx}
\usepackage{epsf}
\usepackage{color}
\usepackage{multirow}

\graphicspath{{Images/}}
\begin{document}

\title{Topological Born-Infeld charged black holes in Einsteinian cubic
gravity}
\author{M. Kord Zangeneh}
\email{mkzangeneh@scu.ac.ir}
\affiliation{Physics Department, Faculty of Science, Shahid Chamran University of Ahvaz,
Ahvaz 61357-43135, Iran}
\author{A. Kazemi}
\email{kazemi.gravity@gmail.com}
\affiliation{Physics Department, Faculty of Science, Shahid Chamran University of Ahvaz,
Ahvaz 61357-43135, Iran}

\begin{abstract}
In this paper, we study four-dimensional topological black hole solutions of
Einsteinian cubic gravity in the presence of nonlinear Born-Infeld
electrodynamics and a bare cosmological constant. First, we obtain the field
equations which govern our solutions. Employing Abbott-Deser-Tekin and Gauss
formulas, we present the expressions of conserved quantities, namely total
mass and total charge of our topological black solutions. We disclose the
conditions under which the model is unitary and perturbatively free of ghosts with
asymptotically (A)dS and flat solutions. We find that, for vanishing bare
cosmological constant, the model is unitary just for asymptotically flat
solutions, which only allow horizons with spherical topology.
We compute the temperature for these solutions and show that it always has a maximum value, which decreases as the values of charge, nonlinear coupling or cubic coupling grows.
Next, we calculate the entropy and electric
potential. We show that the first law of thermodynamics is satisfied for
spherical asymptotically flat solutions. Finally, we peruse the effects of
model parameters on thermal stability of these solutions in both canonical
and grand canonical ensembles.
\end{abstract}

\maketitle

\section{\label{intro}introduction}

General relativity has passed all tests successfully. The latest one was the
detection of gravitational wave \cite{1602.03837}, almost one hundred years
after Einstein predicted it. Despite these achievements, it is unavoidable
to modify general relativity when spacetime curvature becomes extremely large, say, near a singularity.
The most natural modification is to take into account the
higher-order curvature terms. The well-known higher-order Lovelock terms
provide this kind of modification while they respect the constraints of
early version of general relativity \cite{Lovelock1,Lovelock2}. However, these terms
have no contribution in four dimensions. Recently, a cubic order
curvature model with contribution in four dimensions called Einsteinian
cubic gravity (ECG) has been proposed \cite{1607.06463}. This model has attracted a lot of attention \cite{1609.02290, 1610.06675,
1610.08019, 1703.01631, 1703.04625, 1704.02967, 1801.03223, 1802.00018,
1808.01671,1901.03349, 1903.10907, 1909.07983, 1910.10721, 1910.11618,
2002.04071}. ECG respects some of the constraints that Lovelock theories have. For example, on
a maximally symmetric background, it just propagates a transverse and
massless graviton and in all dimensions it has the same relative
coefficients of the different curvature invariants involved. The nontrivial contribution in four dimensions along with other features have made this
model intriguing and important. The latter property allows us to see the
effects of higher-order curvature modifications on ($2+1$)-dimensional
holographic duals of gravity theory solutions.

The solutions in the context of ECG have been explored from different points
of view. In \cite{1609.02290}, by constructing perturbative five-dimensional
black hole solutions of ECG, the holographic entanglement R\'enyi entropy has
been computed in the dual field theory. The first examples of black hole
solutions in ECG have been obtained in \cite{1610.06675}, and thermal
behaviors of them have been explored. In \cite{1610.08019}, the static and
spherically symmetric generalizations of four-dimensional linearly charged
and uncharged black hole solutions in ECG have been constructed and
their thermodynamics has been studied. The most general theory of gravity
to cubic order in curvature called Generalized Quasi-Topological Gravity
(GQTG) whose static spherically symmetric vacuum solutions are fully
described by a single field equation has been constructed in \cite%
{1703.01631}. In this theory, the ECG as well as Lovelock and quasi-topological
gravities have been recovered in four dimensions as special cases. General
results corresponding to static and spherically symmetric black hole
solutions of general higher-derivative gravities including GQTG have also
been established \cite{1703.04625}. It has been proved as well that the
four-dimensional black hole solutions corresponding to an infinite family of
ghost-free higher-order theories are universally stable below a certain mass 
\cite{1704.02967}. In \cite{1801.03223}, by employing the continued fraction
approximation, some interesting properties of ECG black hole solutions such
as the innermost stable circular orbit of massive test bodies near a black
hole and the shadow of a black hole have been obtained. Some properties of
a nonsupersymmetric conformal field theory in three dimensions which could
be a holographic dual to four-dimensional ECG model has been explored in 
\cite{1802.00018}. Euclidean AdS-Taub-NUT and bolt solutions with various
base spaces in four- and six-dimensions constructed respectively in the
context of ECG and GQTG have been studied as well and their thermodynamics features have been explored \cite{1808.01671}.

In this paper, we study four-dimensional topological black hole solutions of
ECG in the presence of nonlinear Born-Infeld (BI) electrodynamics and a bare
cosmological constant. To the best of our knowledge, this is the first consideration of
nonlinearly charged topological solutions in ECG. The importance of
considering nonlinear BI electrodynamics is at least two fold. On the one
hand, it resolves the singularity problem of Maxwell electrodynamics at the
place of point charge \cite{born1934foundations}. On the other hand, it
comes from the low energy limit of open superstring theory \cite%
{fradkin1985effective,metsaev1987born,bergshoeff1987born}. In addition,
photon-photon interaction experiments have suggested that there is a
nonlinear theory of electrodynamics in vacuum \cite{nonlin}. Black holes
with different horizon's topologies show drastically different
thermodynamical properties as well. For instance, whereas Schwarzschild
black holes with spherical horizon are not thermally stable, it has been
argued that Schwarzschild-AdS black holes whose horizons have either planar or hyperbolic topologies are thermally stable and do not underlie Hawking-Page
phase transition \cite{Birmingham1999}. From holographic point of view,
topological solutions are described as duals to thermal states of conformal
field theories as well \cite{top}.

This paper proceeds as follows. In the next section, we will introduce the
action of theory and its field equations. Then, we will calculate conserved
quantities in section \ref{consqu}. In section \ref{sec2}, we will compute thermodynamical quantities and check the satisfaction of
thermodynamics first law. We will explore thermal stability of our solutions
in section \ref{sec3}. Last section is devoted to summary and concluding
remarks.

\section{Action and field equations \label{sec1}}

The Einsteinian cubic gravity (ECG), which is the most general
dimension-independent gravity theory that consists of metric and Riemann
tensor contractions for which linearized spectrum coincides with Einstein
gravity one, in the presence of nonlinear electrodynamics could be written
as \cite{1607.06463,1610.08019}%
\begin{equation}
\mathcal{S}=\frac{1}{16\pi }\int_{\mathcal{M}}d^{4}x\sqrt{-g}\bigg(%
R+\sum_{i=2}^{3}\alpha _{i}L_{i}-2\Lambda _{0}-\lambda \mathcal{P}+\mathcal{L%
}\left( F\right) \bigg),  \label{Action}
\end{equation}%
up to cubic order in curvature. We write action (\ref{Action}) in Planck
units where we set $G=c=1$. In the action above, $\Lambda _{0}$, $\lambda $ and $%
\alpha _{i}$'s are bare cosmological constant, cubic coupling constant and
Lovelock coefficients, respectively. We will assume $\lambda \geq 0$
throughout this paper. Also, $L_{i}$'s stand for $i$th-order Lovelock terms \cite{Lovelock1,Lovelock2}.
Note that $L_2$ is topological and $L_3$ vanishes identically in four
dimensions. The additional cubic contribution $\mathcal{P}$
is defined as \cite{1607.06463}%
\begin{align}
\mathcal{P}=& 12R_{a\text{ }b}^{\ c\text{ }d}R_{c\ d}^{\ e\ f}R_{e\ f}^{\ a\
b}+R_{ab}^{cd}R_{cd}^{ef}R_{ef}^{ab}  \notag \\
& -12R_{abcd}R^{ac}R^{bd}+8R_{a}^{b}R_{b}^{c}R_{c}^{a}\,.  \label{P}
\end{align}%
In this paper, we intend to consider Born-Infeld (BI) nonlinear
electrodynamics for which%
\begin{equation}
\mathcal{L}\left( F\right) =b^{2}\left( 1-\sqrt{1+\frac{F}{2b^{2}}}\right) ,
\label{biL}
\end{equation}%
where $b$ is a nonlinear parameter and $F=F_{ab}F^{ab}$ in which $%
F_{ab}=2\partial _{\lbrack a}A_{b]}$\ and $A_{a}$\ is the electromagnetic
potential. As $b$ tends to infinity, $\mathcal{L}$ reduces to the linear
Maxwell case i.e. $-F/4$. Whereas the second- and third-order Lovelock terms
are respectively topological and trivial in four-dimensions, the new cubic
term $\mathcal{P}$ has contribution in field equations \cite{1610.08019}.

We make the following ansatz for four-dimensional topological nonlinearly
charged black solutions%
\begin{eqnarray}
ds^{2} &=&-N^{2}(r)f(r)dt^{2}+\frac{dr^{2}}{f(r)}+r^{2}d\Omega _{k}^{2}\,,
\label{ansatz} \\
A &=&h\left( r\right) dt,
\end{eqnarray}%
where 
\begin{equation}
d\Omega _{k}^{2}=\left\{ 
\begin{tabular}{ll}
$d\theta ^{2}+\sin ^{2}(\theta )d\phi ^{2}$\ \ \ \ \  & $k=1$ \\ 
$d\theta ^{2}+d\phi ^{2}$ & $k=0$ \\ 
$d\theta ^{2}+\sinh ^{2}(\theta )d\phi ^{2}$ & $k=-1$%
\end{tabular}%
\right. ,
\end{equation}%
represents a $2$-dimensional hypersurface with constant curvature $2k$ and
area $A_{k}$. $k=1$, $0$ and $-1$ represent two-sphere $S^{2}$, plane $%
\mathbb{R}
^{2}$ and hyperbolic $H^{2}$ topologies for the event horizon, respectively.
Varying the action (\ref{Action}) with respect to $N(r)$, $f(r)$ and $h(r)$,
we are left with three field equations. One of these field equations which
arises from variation with respect to $f(r)$, is satisfied by $N(r)=const$.
Then, we have two field equations
\begin{align}
0& =-r^{4}\mathcal{H}-2r^{2}\left( k-\Lambda _{0}r^{2}-rf^{\prime }-f\right)
\notag \\
& -\frac{12\lambda }{r}(r^{3}ff^{\prime \prime 2}+2kr^{2}ff^{\prime \prime
\prime }-4krff^{\prime \prime }  \notag \\
& +r^{3}ff^{\prime }f^{\prime \prime }f^{\prime \prime \prime
2}+4kff^{\prime }+4rff^{\prime 2}-krf^{\prime 2}  \notag \\
& -4f^{2}f^{\prime }-2rf^{2}\left( rf^{\prime \prime \prime }-2f^{\prime
\prime }\right) ),  \label{EOM1} \\
0& =2h^{\prime 3}-b^{2}(2h^{\prime }+rh^{\prime \prime }),  \label{EOM2}
\end{align}%
where $\mathcal{H}=b^{2}-b^{2}\left( 1-h^{\prime 2}/b^{2}\right) ^{-1/2}$,
which is reduced to $-h^{\prime 2}/2$ for Maxwell case.{\ Note that prime
denotes the derivative with respect to $r$. One could immediately solve the
electrodynamic field equation (\ref{EOM2})\ as}

\begin{equation}
h(r)=-\frac{q}{r}\,\mathbf{F}\left( \frac{1}{2},\frac{1}{4},\frac{5}{4},-%
\frac{q^{2}}{b^{2}r^{4}}\right) ,  \label{h(r)}
\end{equation}%
where $\mathbf{F}\left( x,y,z,w\right) $ is the Gauss hypergeometric
function and $q$ is an integration constant related to total electric charge
of black hole. Expanding $h(r)$ about infinity $b$, the potential of Maxwell
electrodynamics can be reproduced as $h(r)=-q/r+\mathcal{O}\left(
b^{-1}\right) $. Substituting $h(r)$ from Eq.~(\ref{h(r)}) to Einstein field
equation~(\ref{EOM1}) and then dividing expressions by $2r^{2}$, one can
integrate Einstein field equation once. After some manipulation, the result
could be simplified as follows 
\begin{align}
& kr-m-\frac{r^{3}\Lambda _{0}}{3}-rf+\frac{1}{6}b^{2}r^{3}  \notag \\
& \times \left[ 1-\,\mathbf{F}\left( -\frac{1}{2},-\frac{3}{4},\frac{1}{4},-%
\frac{q^{2}}{b^{2}r^{4}}\right) \right]  \notag \\
& +\frac{\lambda }{r^{2}}\Big[6rff^{\prime \prime }\left( 2k+rf^{\prime
}-2f\right)  \notag \\
& -2rf^{\prime 2}\left( 3k+rf^{\prime }\right) -12f^{\prime }f\left(
k-f\right) \Big]=0,  \label{eq}
\end{align}%
where $m$ is an integration constant related to total mass of black hole.
For BI-(A)dS gravity in four dimensions ($\lambda =0$), $f(r)$ could be
obtained from above equation as \cite{adsbi1}%
\begin{equation}
f(r)=k-\frac{m}{r}-\frac{r^{2}\Lambda _{0}}{3}+\frac{1}{6}b^{2}r^{2}\left[
1-\,\mathbf{F}\left( -\frac{1}{2},-\frac{3}{4},\frac{1}{4},-\frac{q^{2}}{%
b^{2}r^{4}}\right) \right] . \label{fmet}
\end{equation}%
In this case, the total mass per unit area $A_{k}$ is \cite{adsbi1,adsbi}%
\begin{equation}
M_{E}=\frac{m}{8\pi }.  \label{eimass}
\end{equation}%
In the next section, we turn to calculate the conserved quantities related to
our solutions, namely total mass and total charge.

\section{Conserved quantities \label{consqu}}

In the present section, we intend to compute the total mass and the total charge of our
black solutions. In order to find the total mass, we follow the
Abbott-Deser-Tekin method \cite{adt1} of finding conserved quantities for
higher-order curvature gravities \cite{adt2}. According to which, we have to
find an equivalent quadratic curvature action of the form%
\begin{eqnarray}
\mathcal{S}_{\text{EQCA}}=\int d^{4}x\sqrt{-g}\,\left[ \tilde{\kappa}%
^{-1}\left( R-2\tilde{\Lambda}_{0}\right) +\tilde{\alpha}R^{2}\right. ~~ && 
\notag \\
\left. +\tilde{\beta}R_{ab}R^{ab}+\tilde{\gamma}L_{2}\right] , &&
\label{eqca}
\end{eqnarray}%
where $L_{2}=R_{abcd}R^{abcd}-4R_{ab}R^{ab}+R^{2}$ is the second-order
Lovelock term known as the Gauss-Bonnet term. Note that equivalent quadratic
curvature action has the same vacuum solution and the same linearized field
equations as the corresponding higher-order gravity theory. We can write the
gravity part of our action (\ref{Action}) which is cubic-order as 
\begin{gather}
\mathcal{S}_{G}=\int_{\mathcal{M}}d^{4}x\sqrt{-g}f\left( R_{cd}^{ab}\right) ,
\notag \\
f\left( R_{cd}^{ab}\right) =\kappa ^{-1}\left[ R+\alpha _{2}L_{2}-2{\Lambda
_{0}}-\lambda \mathcal{P}\right] ,  \label{f}
\end{gather}%
in which additional cubic contribution $\mathcal{P}$ defined in Eq. (\ref{P}%
) can be re-written as 
\begin{align}
\mathcal{P}=&
12R_{cd}^{ab}R_{af}^{ce}R_{be}^{df}+4R_{cd}^{ab}R_{ab}^{ef}R_{ef}^{cd} 
\notag \\
& -12R_{cd}^{ab}R_{a}^{c}R_{b}^{d}+8R_{b}^{a}R_{a}^{c}R_{c}^{b}\,.
\label{P1}
\end{align}%
In appendix \ref{ap1}, we show how one can obtain Eq. (\ref{P1}). We suppose
that action (\ref{f}) has an asymptotically (A)dS solution around $%
\bar{g}_{ab}$ for which%
\begin{equation}
\bar{R}_{cd}^{ab}=\frac{\Lambda }{3}\left( \delta _{c}^{a}\delta
_{d}^{b}-\delta _{d}^{a}\delta _{c}^{b}\right),  \label{adsback}
\end{equation}%
where $\Lambda $ is the effective cosmological constant. Indeed, for $%
\Lambda =0$, we have an asymptotically flat solution. Comparing Eq. (\ref{f}), with
the most general cubic curvature action constructed from Riemann tensor
contractions \cite{riempol} 
\begin{eqnarray}
\mathcal{S}_{\text{GCCA}}=\int d^{4}x\sqrt{-g}\,\left[ \kappa ^{-1}\left(
R-2\Lambda _{0}\right) +\alpha R^{2}+\beta R_{ab}R^{ab}\right. ~~ &&  \notag
\\
\left. +\gamma L_{2}+F\left( R_{cd}^{ab}\right) \right] , &&  \notag \\
&&  \label{Lag}
\end{eqnarray}%
in which%
\begin{eqnarray*}
F\left( R_{cd}^{ab}\right) &\equiv
&a_{1}R_{cd}^{ab}R_{af}^{ce}R_{be}^{df}+a_{2}R_{cd}^{ab}R_{ab}^{ef}R_{ef}^{cd}+a_{3}R_{b}^{a}R_{ea}^{cd}R_{cd}^{eb}
\\
&&+a_{4}RR_{cd}^{ab}R_{ab}^{cd}+a_{5}R_{b}^{a}R_{d}^{c}R_{ac}^{bd}+a_{6}R_{b}^{a}R_{a}^{c}R_{c}^{b}
\\
&&+a_{7}RR_{b}^{a}R_{a}^{b}+a_{8}R^{3},
\end{eqnarray*}%
one can find that 
\begin{gather}
\gamma =\frac{\alpha _{2}}{\kappa },\text{ \ }a_{1}=-\frac{12\lambda }{%
\kappa },\text{ \ \ }  \notag \\
\text{ \ }a_{2}=-\frac{4\lambda }{\kappa },\text{ \ }a_{5}=\frac{12\lambda }{%
\kappa },\text{ \ }a_{6}=-\frac{8\lambda }{\kappa },  \notag \\
\alpha =\beta =a_{3}=a_{4}=a_{7}=a_{8}=0.  \label{par}
\end{gather}%
In addition, parameters in actions (\ref{eqca}) and (\ref{Lag}) are related
to each other, so that, in four dimensions \cite{adt2}%
\begin{align}
\frac{1}{\tilde{\kappa}}\equiv & \frac{1}{\kappa }-\frac{\Lambda ^{2}}{3}%
\left[ a_{1}+4a_{2}+6\left( a_{3}+4a_{4}\right) \right.  \notag \\
& \left. \text{ \ \ \ \ \ \ \ }+9\left( a_{5}+a_{6}+4a_{7}+16a_{8}\right) %
\right] , \\
\tilde{\Lambda}_{0}\equiv & \frac{\tilde{\kappa}}{\kappa }\Lambda _{0}+\frac{%
2\Lambda }{3}\left( 1-\frac{\tilde{\kappa}}{\kappa }\right) , \\
\tilde{\alpha}\equiv & \alpha +\frac{\Lambda }{3}\left[
3a_{1}-6a_{2}-8a_{4}+a_{5}\right.  \notag \\
& \left. \text{ \ \ \ \ \ \ \ \ \ \ \ }+3\left( -a_{3}+2a_{7}+12a_{8}\right) %
\right] , \\
\tilde{\beta}\equiv & \beta +\frac{\Lambda }{3}\left[
-9a_{1}+24a_{2}+16a_{3}+5a_{5}\right.  \notag \\
& \left. \text{ \ \ \ \ \ \ \ \ \ \ \ \ }+3\left(
16a_{4}+3a_{6}+4a_{7}\right) \right] , \\
\tilde{\gamma}\equiv & \gamma +\frac{\Lambda }{3}\left[ -3a_{1}+6a_{2}+3%
\left( a_{3}+4a_{4}\right) \right] .
\end{align}%
Using Eq. (\ref{par}), one obtains
\begin{gather}
\frac{\kappa }{\tilde{\kappa}}=1-\frac{8\lambda \Lambda ^{2}}{3},
\label{kap} \\
\Lambda _{0}=\tilde{\Lambda}_{0}\left( 1-\frac{8\lambda \Lambda ^{2}}{3}%
\right) +\frac{16\lambda \Lambda ^{3}}{9},  \label{lam} \\
\kappa \tilde{\gamma}=\left( \alpha _{2}+4\lambda \Lambda \right) , \\
\tilde{\alpha}=\tilde{\beta}=0.  \label{al}
\end{gather}%
Moreover, in four dimensions, we have $\Lambda =\tilde{\Lambda}_{0}$ \cite%
{adt2}. Therefore, Eq. (\ref{lam}) leads to 
\begin{equation}
\frac{8\lambda }{9}\Lambda ^{3}-\Lambda +\Lambda _{0}=0,  \label{lameq}
\end{equation}%
whose solution is the effective cosmological constant $\Lambda $. As Eq. (%
\ref{lameq}) shows, $\Lambda =\Lambda _{0}$ for BI-(A)dS gravity ($\lambda
=0 $). One also needs to calculate the equivalent effective Newton's
constant as%
\begin{equation}
\frac{1}{\tilde{\kappa}_{e}}\equiv \frac{1}{\tilde{\kappa}}+8\Lambda \tilde{%
\alpha}+\frac{4\Lambda }{3}\tilde{\beta}.  \label{eq:ee-Newton}
\end{equation}%
Putting $\tilde{\alpha}=\tilde{\beta}=0$ from Eq. (\ref{al}) to above
relation, we find 
\begin{equation}
\tilde{\kappa}_{e}=\tilde{\kappa}.  \label{kap1}
\end{equation}%
The total mass per unit area $A_{k}$ is given by \cite{adt2}%
\begin{equation}
M=\frac{\kappa }{\tilde{\kappa}_{e}}M_{E},
\end{equation}%
where $M_{E}$ is the total mass in Einstein gravity which can be found as $%
M_{E}=m/8\pi $ in four dimensions (see Eq. (\ref{eimass})). Thus, using Eqs.
(\ref{kap}) and (\ref{kap1}), we have%
\begin{equation}
\frac{\kappa }{\tilde{\kappa}_{e}}=\frac{\kappa }{\tilde{\kappa}}=1-\frac{%
8\lambda \Lambda ^{2}}{3},  \label{kap2}
\end{equation}%
which leads to%
\begin{equation}
M=\left( 1-\frac{8\lambda \Lambda ^{2}}{3}\right) \frac{m}{8\pi }.
\label{hicumass}
\end{equation}%
For model to be unitary and free of ghost, it is necessary to have $\kappa /%
\tilde{\kappa}_{e}>0$ \cite{adt2}. Here, it is remarkable to discuss about
the restricts imposed by Eq. (\ref{lameq}) as well. This equation is cubic
in $\Lambda $ and its discriminant $\Delta $ is%
\begin{equation*}
\Delta =\frac{32\lambda }{9}\left( 1-6\lambda \Lambda _{0}^{2}\right) .
\end{equation*}%
So, it has three real roots if%
\begin{equation*}
\Delta \geq 0\rightarrow \lambda \left( 1-6\lambda \Lambda _{0}^{2}\right)
\geq 0.
\end{equation*}%
One of the possibilities is $\Lambda _{0}=0$ for which%
\begin{equation*}
\Lambda =0\text{ and }\Lambda =\pm \frac{3}{2\sqrt{2\lambda }}.
\end{equation*}%
Note that we have assumed $\lambda $ to be positive throughout the paper.
For $\Lambda =\pm 3/2\sqrt{2\lambda }$, $\kappa /\tilde{\kappa}_{e}$ given
by Eq. (\ref{kap2}) is negative. As a result, for vanishing bare
cosmological constant $\Lambda _{0}$, we have just a unitary model with
asymptotically flat solution ($\Lambda =0$). Another possibility is $\Lambda
_{0}^{2}=1/6\lambda $ for which $\Delta =0$. In this case, we have%
\begin{eqnarray*}
\text{For }\Lambda _{0} &=&+\frac{1}{\sqrt{6\lambda }}:\text{ }\Lambda =-%
\sqrt{\frac{3}{2\lambda }}\text{ and }\Lambda =+\frac{1}{2}\sqrt{\frac{3}{%
2\lambda }}, \\
\text{For }\Lambda _{0} &=&-\frac{1}{\sqrt{6\lambda }}:\text{ }\Lambda =+%
\sqrt{\frac{3}{2\lambda }}\text{ and }\Lambda =-\frac{1}{2}\sqrt{\frac{3}{%
2\lambda }},
\end{eqnarray*}%
where the second one in both cases is a double root. One could check that
for all these roots $\kappa /\tilde{\kappa}_{e}\leq 0$ and therefore the
model is not unitary. For some other values of parameters including ones for
which $\Delta <0$, one can find some vacuum solutions within a unitary
model as well. Note that for $\Delta <0$ ($1-6\lambda \Lambda _{0}^{2}<0$),
the cubic equation (\ref{lameq}) has two imaginary complex conjugate roots
and just one real root for $\Lambda $. In our thermodynamical studies which
will be presented in next sections, we focus on vanishing bare cosmological
constant case ($\Lambda _{0}=0$). As discussed above, in this case, an
asymptotically flat solution ($\Lambda =0$) is just allowed. So, the mass
per unit area reads%
\begin{equation}
M=\frac{m}{8\pi },  \label{mass}
\end{equation}%
according to Eq. (\ref{hicumass}).

Here, we turn to calculate the total charge of our black solutions via Gauss
law. Using (\ref{h(r)}), one can show that $F_{rt}=h^{\prime }(r)$ is 
\begin{equation}
F_{rt}=\frac{q}{r^{2}\sqrt{1+q^{2}/b^{2}r^{4}}}.  \label{Frt}
\end{equation}%
Then, using Gauss law, the total charge is given by%
\begin{equation}
Q=\frac{\,{1}}{4\pi }\int r^{2}\mathcal{L}_{F}F_{\mu \nu }n^{\mu }u^{\nu }d{%
\Omega }_{k},  \label{chdef}
\end{equation}%
where $\mathcal{L}_{F}=\partial \mathcal{L}/\partial F$ and $\mathcal{L}$ is
the Lagrangian of BI nonlinear electrodynamics expressed in Eq. (\ref{biL}).
Also, $n^{\mu }$ and $u^{\nu }$ are the unit spacelike and timelike normals
to the hypersurface of radius $r$ given as $n^{\mu }=\left( \sqrt{-g_{tt}}%
\right) ^{-1}dt=\left( \sqrt{f(r)}\right) ^{-1}dt$ and $u^{\nu }=\left( 
\sqrt{g_{rr}}\right) ^{-1}dr=\sqrt{f(r)}dr$. Using Eqs. (\ref{Frt}) and (\ref%
{chdef}), we can obtain the total charge of black solutions per unit area as 
\begin{equation}
Q=\frac{q}{16\pi }.  \label{charge}
\end{equation}%
In the next section, we will obtain thermodynamical quantities corresponding to
our solutions and check the first law of thermodynamics for them.

\section{Thermodynamical quantities and Thermodynamics first law \label{sec2}}

In this section, we intend to calculate thermodynamical quantities and check
the first law of thermodynamics. For this purpose, we first have to compute
field equations near the black hole horizon. Taylor expansion of metric
function $f\left( r\right) $ near the black hole horizon $r_{h}$ is%
\begin{equation}
f(r)=\sum_{n=0}^{\infty }a_{n}(r-r_{h})^{n}\,,  \label{expansion}
\end{equation}%
in which $a_{n}=f^{(n)}(r_{h})/n!$. Note that $a_{0}=f(r_{h})=0$ and $%
a_{1}=f^{\prime }(r_{h})=2\kappa _{g}$ where $\kappa _{g}$ is surface
gravity on the horizon and $f^{\prime }(r_{h})\geq 0$. Plugging above
expansion into field equation (\ref{eq}), one receives the following equation up
to quadratic order of $r-r_{h}$: 
\begin{align}
& kr_{h}-m-8\lambda \kappa _{g}^{2}(2\kappa _{g}+\frac{3k}{r_{h}})-\frac{%
\Lambda _{0}}{3}r_{h}^{3}  \notag \\
& +\frac{1}{6}r_{h}^{3}b^{2}\left[ 1-\mathbf{F}\left( -\frac{1}{2},-\frac{3}{%
4},\frac{1}{4},-\frac{q^{2}}{b^{2}r_{h}^{4}}\right) \right] \,+\Bigg[%
k-\Lambda _{0}r_{h}^{2}  \notag \\
& -2\kappa _{g}r_{h}-24\lambda k\frac{\kappa _{g}^{2}}{r_{h}^{2}}+\frac{1}{2}%
r_{h}^{2}b^{2}\left( 1-\sqrt{1+\frac{q^{2}}{b^{2}r_{h}^{4}}}\right) \Bigg]%
(r-r_{h})  \notag \\
& +\Bigg[72a_{3}\lambda \kappa _{g}(\kappa _{g}+\frac{k}{r_{h}}%
)+24a_{2}^{2}\lambda \kappa _{g}-a_{2}r_{h}-2\kappa _{g}-\Lambda _{0}r_{h} 
\notag \\
& -\frac{24a_{2}\lambda \kappa _{g}}{r_{h}}\left( 4\kappa _{g}+\frac{3k}{%
r_{h}}\right) +\frac{1}{2}r_{h}b^{2}\left[ 1-\left( 1+\frac{q^{2}}{%
b^{2}r_{h}^{4}}\right) ^{-\frac{1}{2}}\right]  \notag \\
& +\frac{72\lambda \kappa _{g}^{2}}{r_{h}^{3}}k+\frac{96\lambda \kappa
_{g}^{3}}{r_{h}^{2}}\Bigg](r-r_{h})^{2}+\mathcal{O}((r-r_{h})^{3})=0.
\label{expan}
\end{align}%
Solving the above equation order by order in terms of $r-r_{h}$ powers, up to
second term, we get: 
\begin{align}
& kr_{h}-m-8\lambda \kappa _{g}^{2}\left( 2\kappa _{g}+\frac{3k}{r_{h}}%
\right) -\frac{\Lambda _{0}}{3}r_{h}^{3}  \notag \\
& +\frac{1}{6}b^{2}r_{h}^{3}\left[ 1-\,\mathbf{F}\left( -\frac{1}{2},-\frac{3%
}{4},\frac{1}{4},-\frac{q^{2}}{b^{2}r_{h}^{4}}\right) \right] =0,  \label{M}
\\
& k+\frac{1}{2}r_{h}^{2}b^{2}\left( 1-\sqrt{1+\frac{q^{2}}{b^{2}r_{h}^{4}}}%
\right) -\Lambda _{0}r_{h}^{2}  \notag \\
& -2\kappa _{g}r_{h}-24\lambda \frac{\kappa _{g}^{2}}{r_{h}^{2}}k=0.
\label{kg}
\end{align}%
As expected, these relations reproduce the following ECG-Maxwell results if $%
b\rightarrow \infty $ \cite{1610.08019} 
\begin{align}
kr_{h}-m-8\lambda \kappa _{g}^{2}\left( 2\kappa _{g}+\frac{3k}{r_{h}}\right)
-\frac{\Lambda _{0}}{3}r_{h}^{3}+\frac{q^{2}}{4r_{h}}& =0\,, \\
k-\frac{q^{2}}{4r_{h}^{2}}-\Lambda _{0}r_{h}^{2}-2\kappa _{g}r_{h}-24\lambda 
\frac{\kappa _{g}^{2}}{r_{h}^{2}}k& =0\,.
\end{align}%
It is notable to mention that since the third term in (\ref{expan}) is
linear in terms of $a_{3}$, we can compute that in terms of $a_{2}$.
Also, other higher order terms are linear with respect to other coefficients
and therefore, all of them could finally be determined in terms of $a_{2}$.
Consequently, the family of solutions have only one free parameter, $a_{2}$.
This fact shows that the present model allows black solutions which have
regular horizons where the regularity condition reduces the number of
solutions from a two-parameter family to a one-parameter one. 
\begin{figure*}[t]
\centering%
\subfigure[~$q=1.5$ and $b=1.4$] {\label{fig1a}
\includegraphics[width=.32\textwidth]{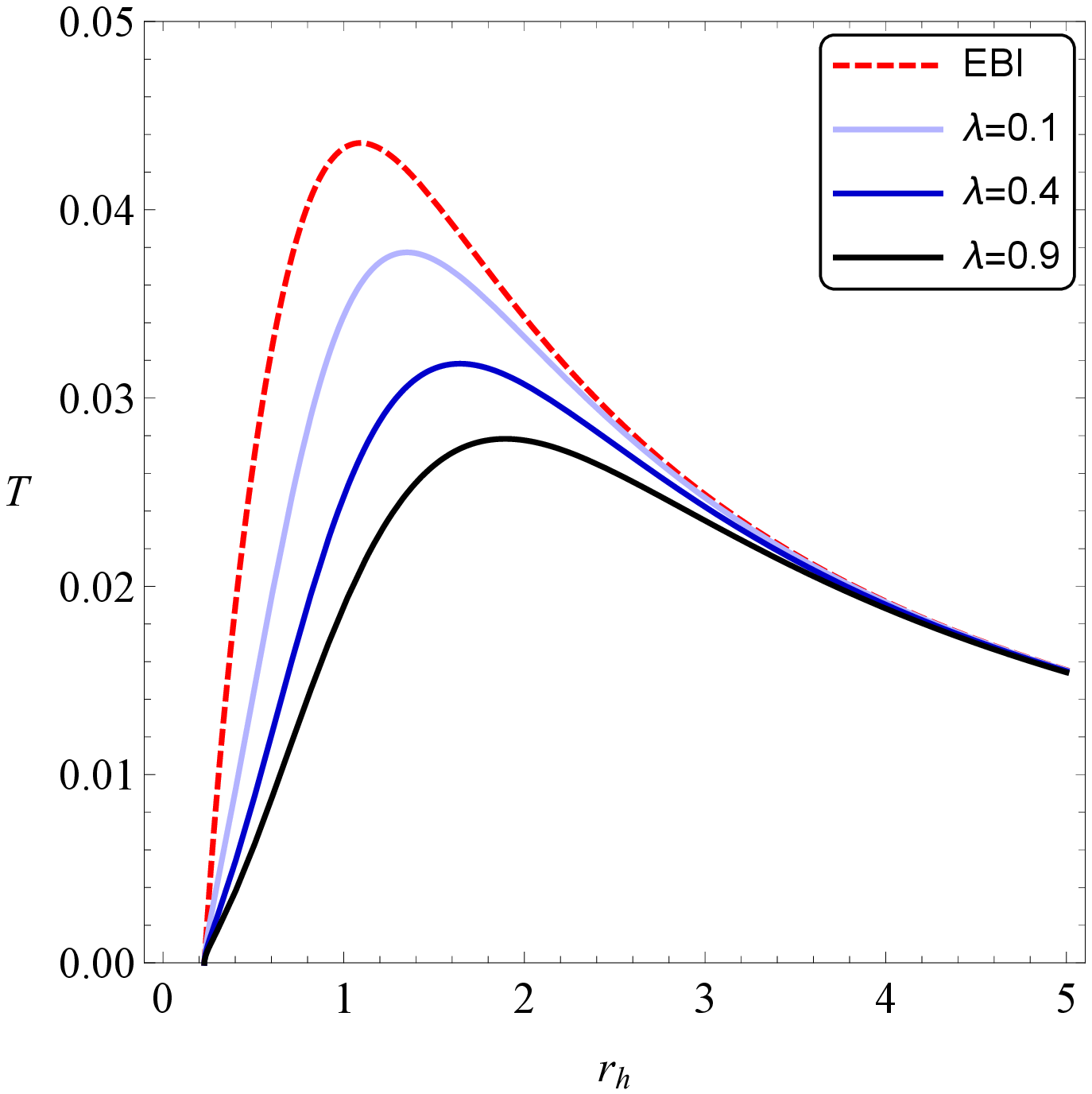}} 
\subfigure[~$q=1.5$ and $\lambda=0.4$] {\label{fig1b}
\includegraphics[width=.32\textwidth]{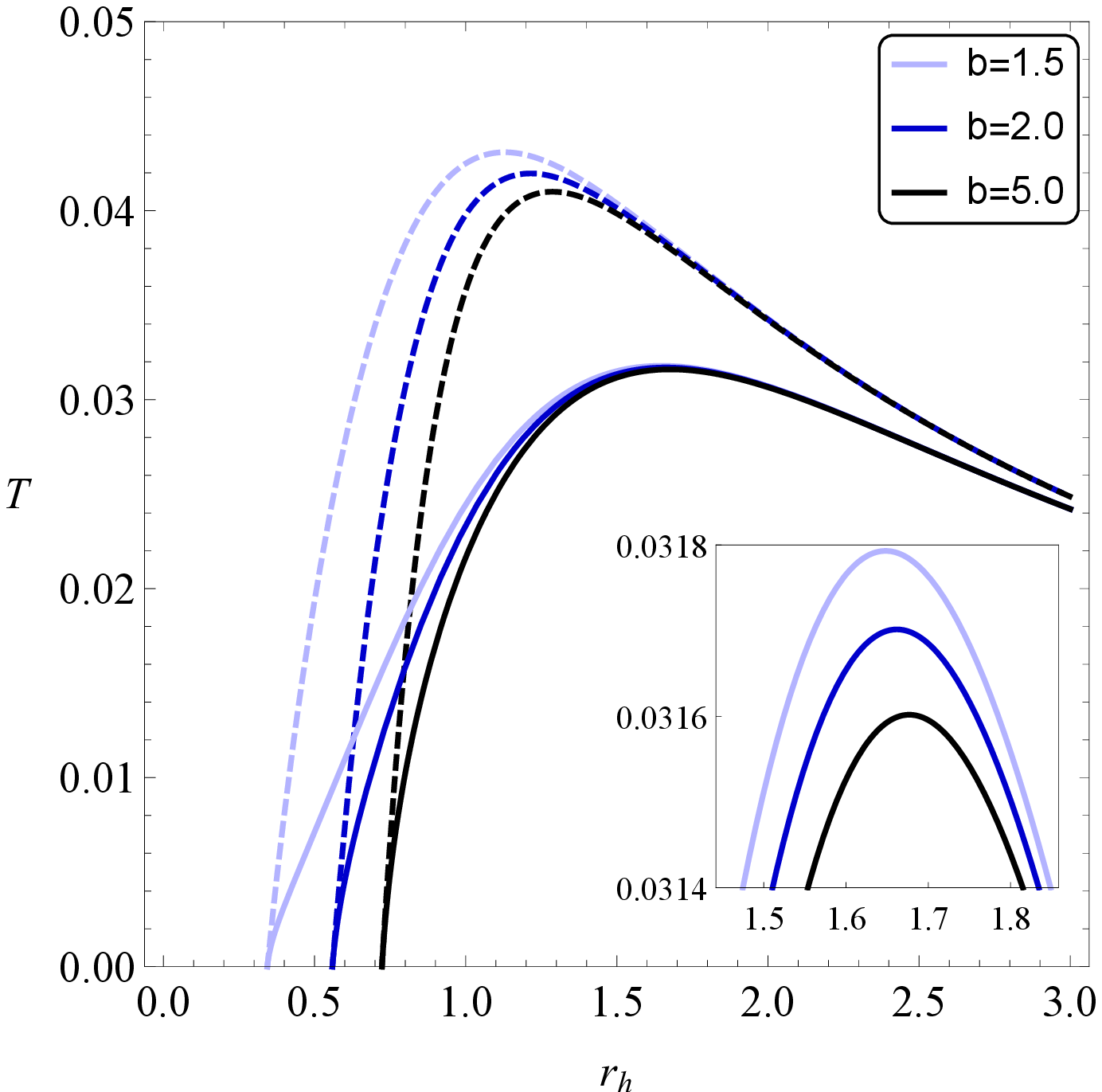}} 
\subfigure[~$b=1.3$ and $\lambda=0.8$] {\label{fig1c}
\includegraphics[width=.32\textwidth]{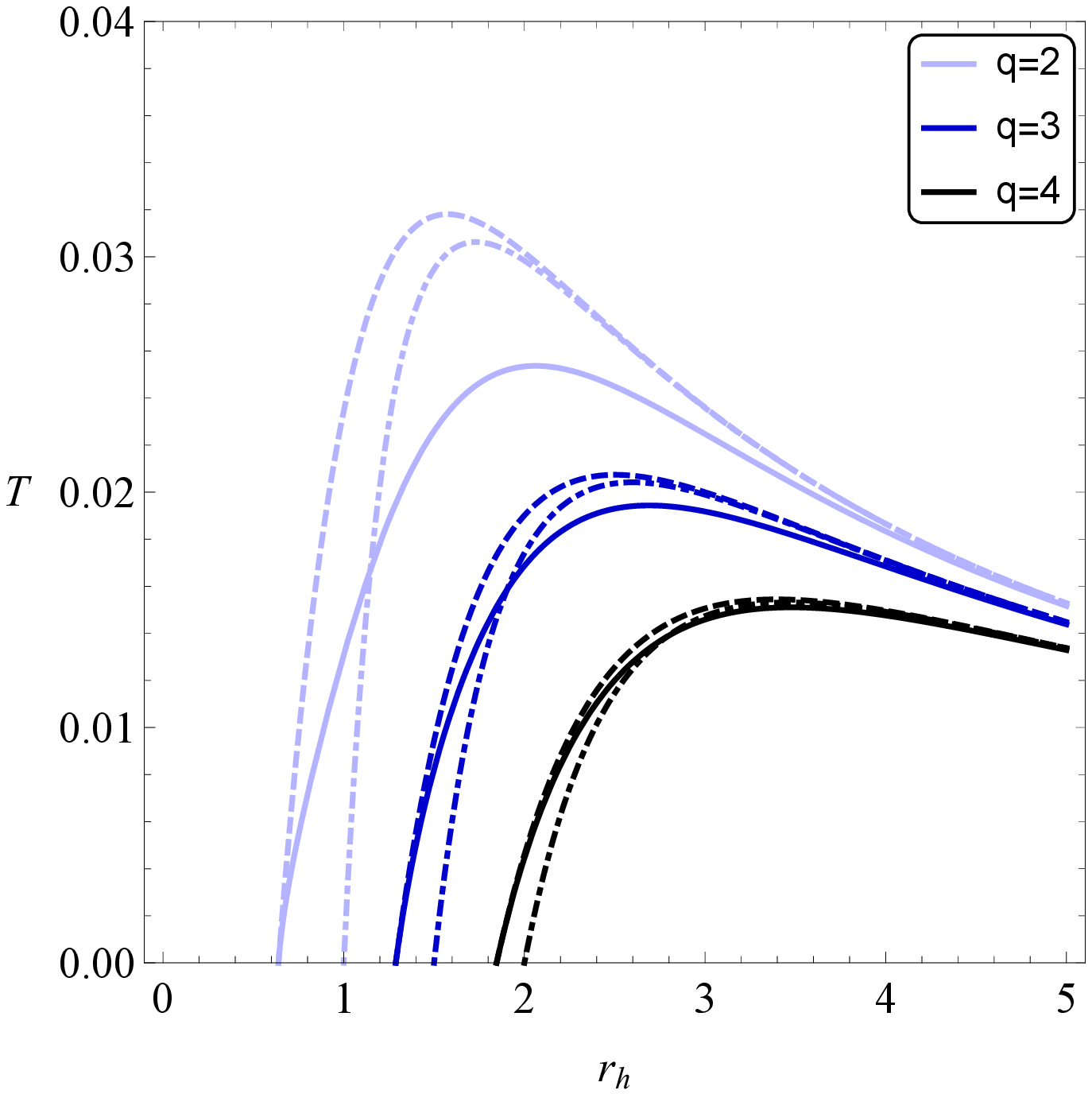}}
\caption{The behavior of $T$ versus $r_{h}$ for spherical asymptotically
flat solutions with various values of $\protect\lambda $, $b$ and $q$. Note
that the solid (dashed) curves show the temperature in the presence
(absence) of cubic terms. In the right panel, dot-dashed curves show the
temperature for Einstein-Maxwell regime.}
\label{fig1}
\end{figure*}

From Eqs. (\ref{M}) and (\ref{kg}), we could determine $m$ and surface gravity $%
\kappa _{g}$ as functions of model parameters:{\ 
\begin{align}
\kappa _{g}=& \frac{r_{h}^{3}}{24\lambda k}\Bigg[1+\frac{12k\lambda b^{2}}{%
r_{h}^{2}}\left( 1-\sqrt{1+\frac{q^{2}}{b^{2}r_{h}^{4}}}\right)  \notag \\
& -\frac{24k\lambda \Lambda _{0}}{r_{h}^{2}}+\frac{24k^{2}\lambda }{r_{h}^{4}%
}\Bigg]^{\frac{1}{2}}-\frac{r_{h}^{3}}{24\lambda k},  \label{kappa_g} \\
\frac{m}{r_{h}}& =k-\frac{24\lambda \kappa _{g}^{2}}{r_{h}^{2}}k-\frac{%
\Lambda _{0}r_{h}^{2}}{3}-\frac{16\lambda \kappa _{g}^{3}}{r_{h}}  \notag \\
& +\frac{b^{2}r_{h}^{2}}{6}\left[ 1-\,\mathbf{F}\left( -\frac{1}{2},-\frac{3%
}{4},\frac{1}{4},-\frac{q^{2}}{b^{2}r_{h}^{4}}\right) \right] .  \label{Mass}
\end{align}%
} It is notable to point out that $m$ is related to total mass $M$ via Eq. (%
\ref{hicumass}). If $\lambda$ and $b$ tends to zero and infinity,
respectively, the above equations reduce to RN-(A)ds black hole ones:{%
\begin{align}
\kappa _{g}& =\frac{k}{2r_{h}}-\frac{q^{2}}{8r_{h}^{3}}-\frac{\Lambda
_{0}r_{h}}{2},  \label{eikg} \\
\frac{m}{r_{h}}& =k-\frac{\Lambda _{0}r_{h}^{2}}{3}+\frac{q^{2}}{4r_{h}^{2}}.
\end{align}%
}The Hawking temperature of our solution can be written in terms of surface
gravity \cite{HawTemp}{%
\begin{equation}
T=\frac{\kappa _{g}}{2\pi },  \label{Temp}
\end{equation}%
where $\kappa _{g}$ has been given in Eq.~(\ref{kappa_g}).} Using (\ref%
{kappa_g}) and (\ref{Temp}), one could find the temperature as%
\begin{align}
T& =\frac{r_{h}^{3}}{48\pi \lambda k}\Bigg[1+\frac{12k\lambda b^{2}}{%
r_{h}^{2}}\left( 1-\sqrt{1+\frac{q^{2}}{b^{2}r_{h}^{4}}}\right)+
\frac{24k^{2}\lambda }{r_{h}^{4}}\Bigg]^{\frac{1}{2}}  \notag \\
& \text{\ \ \ }-\frac{r_{h}^{3}}{48\pi
\lambda k},  \label{Ts}
\end{align}%
where we set $\Lambda _{0}=0$. Note that, according to discussions
presented in section \ref{consqu}, we study thermodynamics of asymptotically
flat solutions ($\Lambda =0$) with vanishing bare cosmological constant $%
\Lambda _{0}$.

Here, it is worthwhile to discuss about the allowed values of $k$ for asymptotically flat solutions.
We know that for Einstein-Maxwell gravity case, the metric function of  asymptotically flat topological solution is $f(r)= k - m/r + q^{2}/r^{2}$, which obviously does not present viable black holes if $k=0$ and $k=-1$ since that would imply a wrong sign for the metric for large values of radial coordinate $r$ where $-m/r$ or $-1$ would dominate. It is exactly the case for bare ECG because the contributions
from the cubic piece are very small, asymptotically (for large $r$) as well. For the BI charged black holes in ECG, this reasoning holds since the term related to nonlinear electrodynamics in metric function (as appears in Eq. (\ref{fmet})) is asymptotically proportional to $r^{-2}$. So, we focus on spherically symmetric black hole solutions ($k=1$) because those are the only ones which could exist in asymptotically flat space. One also could check that in hyperbolic and planar cases, the temperature given by Eq. (\ref{Ts}) is negative and from this point of view, as well, these cases are not physically meaningful. Therefore, we study the thermodynamics of asymptotically flat solutions with spherical topology on horizon.

Now, we are going to discuss the effects of each model parameter $\lambda $%
, $b$ and $q$ on the temperature of spherical asymptotically flat solutions.
From Eq. (\ref{Ts}), one can see that for $k=1$, the first term of temperature formula is positive
while the second term is negative. Increasing parameters $b$ and $q$
enhances the magnitude of
\begin{equation*}
\frac{12k\lambda b^{2}}{r_{h}^{2}}\left( 1-\sqrt{1+\frac{q^{2}}{%
b^{2}r_{h}^{4}}}\right) ,
\end{equation*}%
in Eq. (\ref{Ts}). This term is negative for $k=1$, so, as $b$ or $q$ increases, the temperature $T$ of spherical asymptotically flat solutions decreases.
In order to exhibit these effects, we have plotted $T$ versus $r_{h}$
for different values of $b$, $q$ and $\lambda $ in Fig. \ref{fig1}. This
figure shows that as each one of parameters $b$, $q$ or $\lambda $ grows,
the temperature value becomes lower. This means that as the effect of the cubic term grows (or equivalently, as the nonlinearity of electrodynamics weakens), the temperature of
spherical black holes decreases. Fig. \ref{fig1} shows that for spherical
asymptotically flat solutions, there is a maximum temperature $T_{\text{max}}
$. \ The behavior of $T_{\text{max}}$ in terms of model parameters is
exhibited in table~\ref{table1} as well. One can find that the value of $T_{%
\text{max}}$ decreases as each of the parameters $\lambda $, $b$ and $q$
increases.

\begin{table}[t]
\caption{The behavior of maximum temperature $T_{\text{max}}$ for various
values of model parameters.}
\label{table1}%
\begin{tabular}{cccc}
\hline\hline
$\lambda $ & $b$ & $q$ & $T_{\text{max}}$ \\ \hline
\ \ \ \ \ \ 0.01 \ \ \ \ \  &  &  & \ \ \ \ \ \ 0.10 \ \ \ \ \  \\ 
0.02 & \ \ \ \ \ \ 0.2 \ \ \ \ \  & \ \ \ \ \ \ 0.5 \ \ \ \ \  & 0.08 \\ 
0.05 &  &  & 0.06 \\ \hline
& 0.01 &  & 0.10 \\ 
0.01 & 0.50 & 1 & 0.08 \\ 
& 0.95 &  & 0.07 \\ \hline
&  & 0.8 & 0.08 \\ 
0.01 & 0.6 & 1.2 & 0.07 \\ 
&  & 1.6 & 0.06 \\ \hline\hline
\end{tabular}%
\end{table}

\begin{figure*}[t]
\centering%
\subfigure[~] {\label{fig2a}
\includegraphics[width=.32\textwidth]{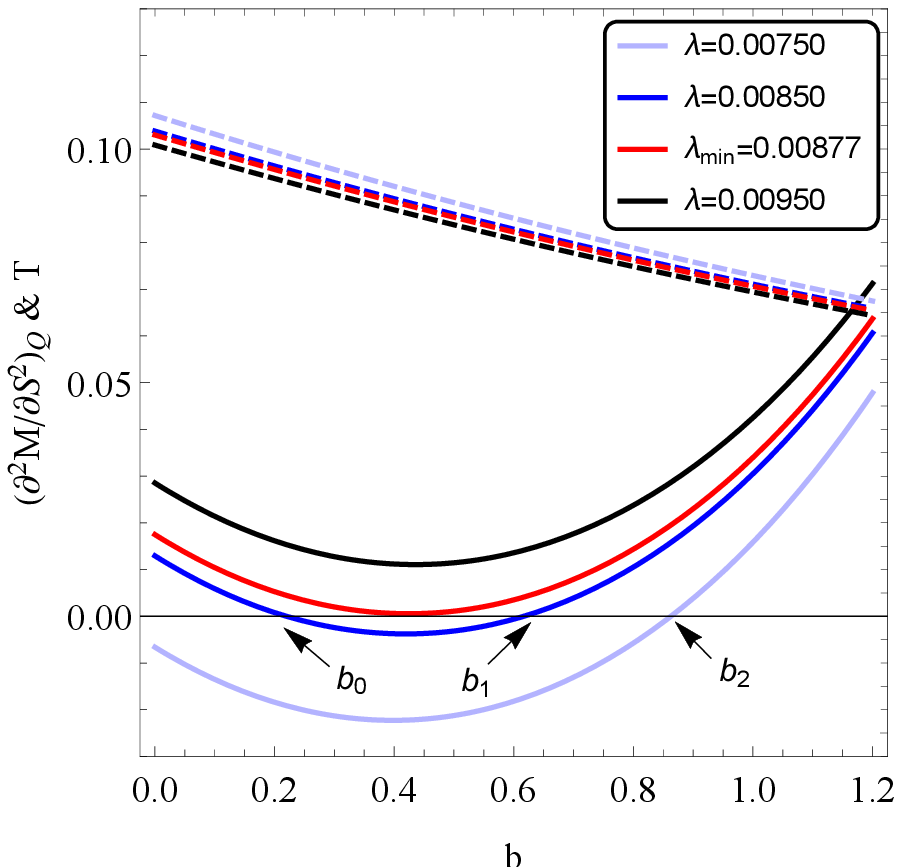}} \qquad \qquad 
\subfigure[~] {\label{fig2b}
	\includegraphics[width=.32\textwidth]{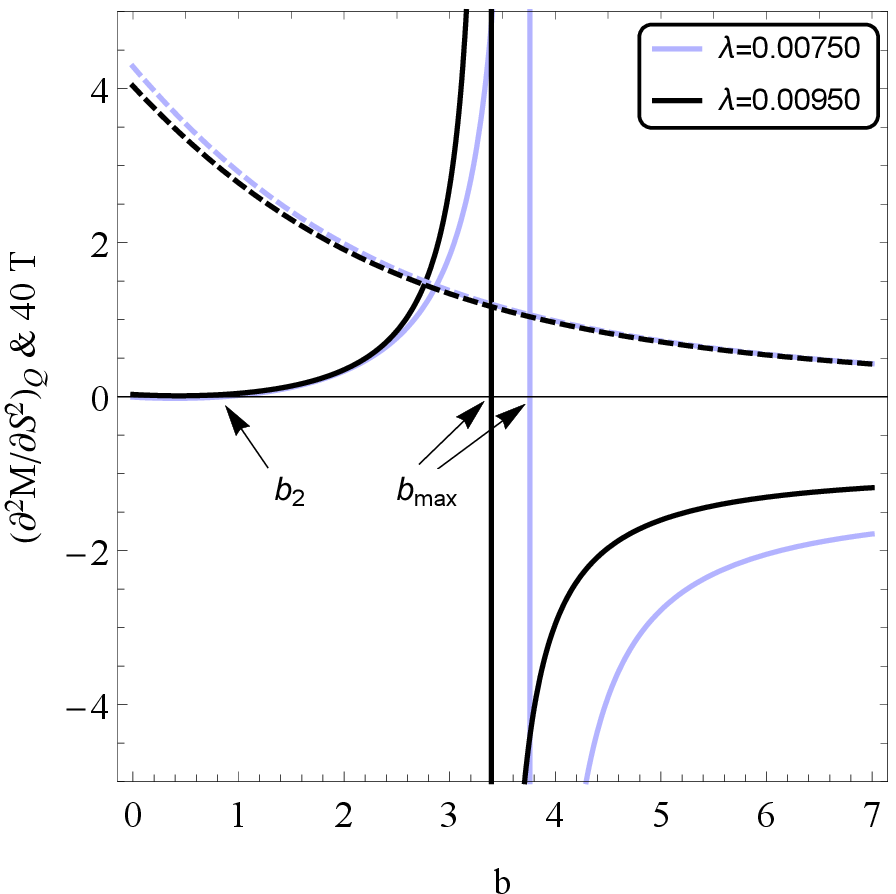}}
\caption{The behaviors of $(\partial ^{2}M/\partial S^{2})_{Q}$ and $T$
(dashed) versus $b$ for different values of $\protect\lambda $ with $q=1$
and $r_{h}=0.5$.}
\label{fig2}
\end{figure*}

Let us now calculate the entropy of our black holes. The entropy of black
solutions in higher-order gravities could be computed by Wald's formula
given by \cite{entropy} 
\begin{equation}
S=-2\pi \int_{H}d^{2}x\sqrt{h}\frac{\delta \mathcal{L}_{G}}{\delta R_{abcd}}%
\epsilon _{ab}\epsilon _{cd}\,,
\end{equation}%
where $h$ is the determinant of the induced metric on the horizon, $\delta 
\mathcal{L}_{G}/\delta R_{abcd}$ is the Euler-Lagrange derivative of
gravitational Lagrangian and $\epsilon _{ab}$ is the binormal of the horizon
normalized as $\epsilon _{ab}\epsilon ^{ab}=-2$. Applying the above formula on
our theory introduced by action (\ref{Action}), we receive
\begin{align}
S=& \frac{1}{4}\int_{H}d^{2}x\sqrt{h}\Big[1+2\alpha _{2}R_{(2)}+\lambda \Big(%
36R_{b\ d}^{\ e\ f}R_{aecf}  \notag \\
& +3R_{ab}^{\ \ ef}R_{cdef}-12R_{ac}R_{db}-24R^{ef}R_{ebfc}g_{bd}  \notag \\
& +24g_{bd}R_{ce}R_{\ a}^{e}\Big)\epsilon ^{ab}\epsilon ^{cd}\Big]\,,
\end{align}%
in which $R_{(2)}$ is the Ricci scalar of the induced metric on the
horizon. This term comes from the Gauss-Bonnet term $%
L_{2}=R^{2}-4R_{ab}R^{ab}+R_{abcd}R^{abcd}$ in the action (\ref{Action}). As we have
pointed out before, this term is topological and hence has no effect in our
calculations so far. However, it contributes to the entropy. Using the
metric (\ref{ansatz}) with $N=1$, one determines the entropy per unit area as
\begin{equation}
S=\frac{r_{h}^{2}}{4}\left[ 1-24\lambda \frac{\kappa _{g}^{2}}{r_{h}^{2}}%
\left( \frac{2k}{\kappa _{g}r_{h}}+1\right) \right] +k\alpha _{2}\,,
\label{Entropy}
\end{equation}%
in which $\kappa _{g}$ could be replaced by Eq. {(\ref{kappa_g}). It is
remarkable to mention that, the above relation reduces to }$r_{h}^{2}/4$ \cite%
{adsbi1} if the higher-order terms disappear ($\lambda =\alpha _{2}=0$).

Now, we turn to calculate the electric potential. The electric potential $U$%
, measured at infinity with respect to horizon is defined by%
\begin{equation}
U=A_{\mu }\chi ^{\mu }\left\vert _{r\rightarrow \infty }-A_{\mu }\chi ^{\mu
}\right\vert _{r=r_{h}},  \label{Pot}
\end{equation}%
where $\chi =\partial _{t}$ is the null generator of the horizon. Using (\ref%
{h(r)}) and (\ref{Pot}), one finds%
\begin{equation}
U=\frac{q}{r_{h}}\,\mathbf{F}\left( \frac{1}{2},\frac{1}{4},\frac{5}{4},-%
\frac{q^{2}}{b^{2}r_{h}^{4}}\right) .  \label{Pott}
\end{equation}%
\begin{figure*}[t]
\centering%
\subfigure[~] {\label{fig3a}
\includegraphics[width=.32\textwidth]{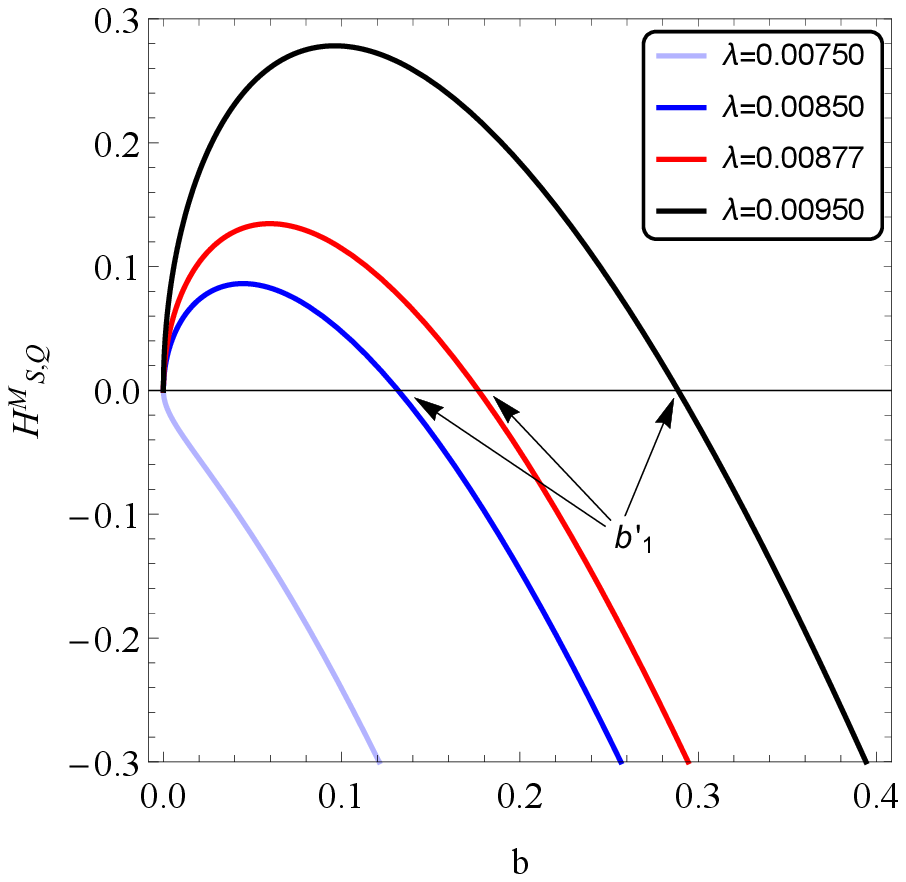}} \qquad \qquad 
\subfigure[~] {\label{fig3b}
	\includegraphics[width=.32\textwidth]{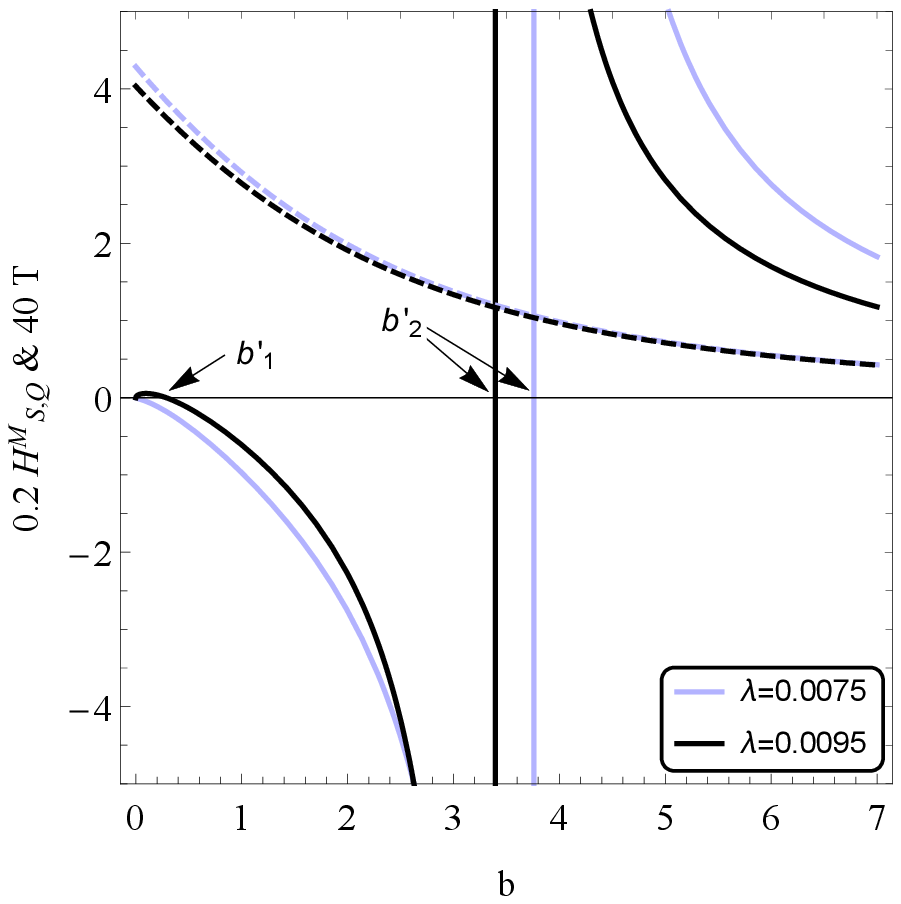}}
\caption{The behavior of $\mathbf{H}_{S,Q}^{M}$ and $T$ (dashed) versus $b$
for different values of $\protect\lambda $ with $q=1$ and $r_{h}=0.5$.}
\label{fig3}
\end{figure*}
\begin{figure*}[t]
\centering%
\subfigure[~$(\partial ^{2}M/\partial S^{2})_{Q}$ and $T$ vs $b$] {\label{fig4a}
\includegraphics[width=.32\textwidth]{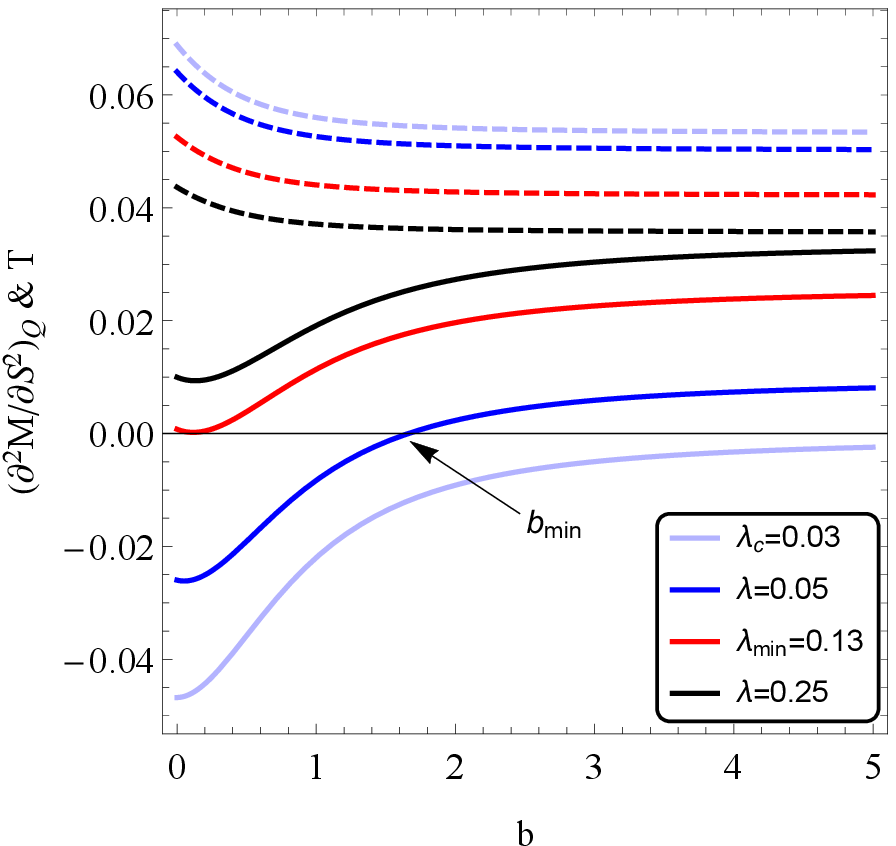}} \qquad \qquad 
\subfigure[~$\mathbf{H}_{S,Q}^{M}$ vs $b$] {\label{fig4b}
	\includegraphics[width=.32\textwidth]{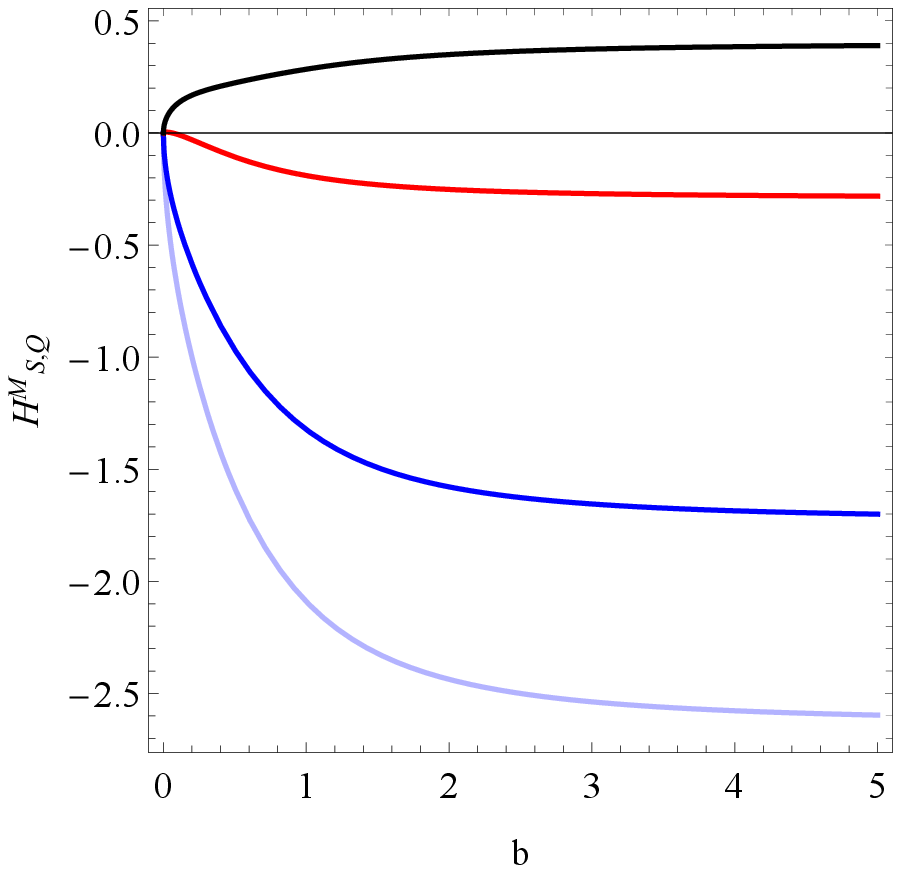}}
\caption{The behaviors of $(\partial ^{2}M/\partial S^{2})_{Q}$, $\mathbf{H}%
_{S,Q}^{M}$ and $T$ (dashed) versus $b$ for different values of $\protect%
\lambda $ with $q=1$ and $r_{h}=1$.}
\label{fig4}
\end{figure*}

In order to check the first law of thermodynamics, we first write a
Smarr-type formula. Using Eqs. (\ref{mass}), (\ref{charge}) and (\ref{Mass}%
), the Smarr-type formula $M\left( r_{h},Q\right) $ can be written as%
\begin{align}
M\left( r_{h},Q\right) & =\frac{1}{8\pi }\Bigg[kr_{h}-\frac{4\lambda \kappa
_{g}^{2}}{r_{h}}\left( 6k+4\kappa _{g}r_{h}\right)  \notag \\
& +\frac{1}{6}b^{2}r_{h}^{3}\left( 1-\,\mathbf{F}\left( -\frac{1}{2},-\frac{3%
}{4},\frac{1}{4},-\frac{\left( 16\pi Q\right) ^{2}}{b^{2}A_{k}^{2}r_{h}^{4}}%
\right) \right) \Bigg].  \label{Mrh}
\end{align}%
According to Eq. (\ref{Entropy}), $r_{h}=r_{h}(S,Q)$ and in general $%
M=M(S,Q) $. We can then consider $S$ and $Q$ as a complete set of extensive
quantities for mass. Therefore, temperature $T$ and electric potential $U$
are defined as conjugate intensive quantities for $S$ and $Q$, respectively
and%
\begin{equation*}
T=\left( \frac{\partial M}{\partial S}\right) _{Q},\text{ \ \ \ \ }U=\left( 
\frac{\partial M}{\partial Q}\right) _{S}.
\end{equation*}%
These intensive quantities can be calculated using Eqs. (\ref{charge}), \ (%
\ref{Entropy}) and (\ref{Mrh}) where\footnote{%
To obtain Eq. (\ref{U0}), we use cyclic rule as $\left( \frac{\partial r_{h}%
}{\partial Q}\right) _{S}=-\frac{\left( \partial S/\partial Q\right) _{rh}}{%
\left( \partial S/\partial r_{h}\right) _{Q}}$.}%
\begin{eqnarray}
T &=&\left( \frac{\partial M}{\partial S}\right) _{Q}=\left( \frac{\partial M%
}{\partial r_{h}}\right) _{Q}\left( \frac{\partial S}{\partial r_{h}}\right)
_{Q}^{-1},  \label{T0} \\
U &=&\left( \frac{\partial M}{\partial Q}\right) _{S}=\left( \frac{\partial M%
}{\partial Q}\right) _{r_{h}}+\left( \frac{\partial M}{\partial r_{h}}%
\right) _{Q}\left( \frac{\partial r_{h}}{\partial Q}\right) _{S}  \notag \\
&=&\left( \frac{\partial M}{\partial Q}\right) _{r_{h}}-\frac{\left( \frac{%
\partial S}{\partial Q}\right) _{r_{h}}\left( \frac{\partial M}{\partial
r_{h}}\right) _{Q}}{\left( \frac{\partial S}{\partial r_{h}}\right) _{Q}} 
\notag \\
&=&\left( \frac{\partial M}{\partial Q}\right) _{r_{h}}-T\left( \frac{%
\partial S}{\partial Q}\right) _{r_{h}}.  \label{U0}
\end{eqnarray}%
Our calculations show that $T$ and $U$ computed by Eqs. (\ref{T0}) and (\ref%
{U0}) coincide with ones obtained from Eqs. (\ref{Ts}) and (\ref{Pott}) for
nonlinearly charged spherical ($k=1$) asymptotically flat solutions. Thus,
the first law of thermodynamics%
\begin{equation}
dM=TdS+UdQ,  \label{TFL}
\end{equation}%
is satisfied by these quantities.

In next section, we will study the thermal stability of spherical solutions
in both canonical and grand canonical ensembles.

\section{Thermal stability \label{sec3}}

In this section, we intend to study thermal stability of the
four-dimensional nonlinearly charged asymptotically flat solutions of ECG
with spherical horizon in both canonical and grand canonical ensembles. In
canonical ensemble where the charge is a fixed parameter, the positivity of
heat capacity $C=T/(\partial T/\partial S)_{Q}=T/(\partial ^{2}M/\partial
S^{2})_{Q}$ guarantees the local stability \cite{stab}. Therefore, it is
sufficient to check that $(\partial ^{2}M/\partial S^{2})_{Q}$ is positive
in order to explore the stability of solutions in the ranges where
temperature is positive as well. Charge $Q$ is no longer fixed in the grand
canonical ensemble. In this ensemble, the system is locally stable
provided that the determinant of Hessian matrix $\mathbf{H}_{S,Q}^{M}=\left[
\partial ^{2}M/\partial S\partial Q\right] $ is positive, in positive
temperature ranges. Since the explicit forms of $(\partial ^{2}M/\partial
S^{2})_{Q}$ and $\mathbf{H}_{S,Q}^{M}$ are complicated, we avoid writing
them here. However, the main results are depicted in Figs.~\ref{fig2}-\ref%
{fig4}.

Let us now turn to study the stability of black hole solutions in canonical and
grand canonical ensembles. Our investigations show that the values of charge 
$q$ and horizon radius $r_{h}$ determine how thermal stability is influenced
by the other parameters. We study thermal stability in canonical ensemble for
two different types of solutions (Figs. \ref{fig2} and \ref{fig4a}). Let us
focus on the first type exhibited in Fig. \ref{fig2}. As Fig. \ref{fig2a}
shows, in highly nonlinear electrodynamics regime (small nonlinearity
parameter $b$ values), there is a minimum value for cubic coupling $\lambda $
($\lambda _{\min }$) that for values greater than it, the system is stable
up to $b_{\max }$ (Fig. \ref{fig2b}). For $\lambda $ values a bit smaller
than $\lambda _{\min }$, the system is stable except for $b$ values between $%
b_{0}$ and $b_{1}$ (Fig. \ref{fig2a}). As $\lambda $ becomes smaller, $b_{0}$
tends to zero and $b_{1}$ becomes greater. So, the instability interval
becomes larger. In this case, the system is stable for $b$ values between $%
b_{2}$ and $b_{\max }$ (Fig. \ref{fig2b}). $b_{2}$ and $b_{\max }$ increase
as $\lambda $ decreases. Also, for $b$ values greater than $b_{\max }$
(including linear Maxwell case), the system is unstable. Note that we plot
the curves in Fig. \ref{fig2b} just for two $\lambda $ values\ in order not
to have a messy plot. For other $\lambda $ values, the qualitative behavior
is the same. In addition, we plot the temperature in Figs. \ref{fig2}-\ref%
{fig4} as well, to ensure it is positive in the regions under study. One
could see that the behavior of temperature in terms of $\lambda $ and $b$
coincides with what we have discussed before i.e. temperature decreases as each of the
latter parameters increases. In second type displayed in Fig. \ref{fig4a},
we again have a minimum value for $\lambda $. However, in this case, there
is no upper bound for $b$ and the system is stable for all $b$ values if $%
\lambda \geq \lambda _{\min }$. For $\lambda $ values a bit smaller than $%
\lambda _{\min }$, there is a minimum value for $b$ ($b_{\min }$) as well,
that for $b>b_{\min }$, we have a stable system. As $\lambda $ becomes
smaller, we have a critical value $\lambda _{c}$ that for $\lambda <\lambda
_{c}$, the system is unstable for all $b$ values.

We continue by discussing the thermal stability in grand canonical ensemble.
For the first type represented in Fig. \ref{fig3}, in highly nonlinear
electrodynamics regime (Fig. \ref{fig3a}), the system is unstable for some
cubic couplings, however for some greater ones, the system becomes stable up
to a specific value of $b$ ($b_{1}^{\prime }$). For $b$ values greater than $%
b_{1}^{\prime }$, we have an unstable system between $b_{1}^{\prime }$ and $%
b_{2}^{\prime }$ (Fig. \ref{fig3b}). For $b>b_{2}^{\prime }$ (including
linear Maxwell case), the system becomes thermally stable. $b_{1}^{\prime }$
($b_{2}^{\prime }$) is greater (smaller) for greater values of $\lambda $.
In second type displayed in Fig. \ref{fig4b}, whereas the system is unstable
for some values of $\lambda $, it becomes totally stable as $\lambda $ grows.

\section{Summary and concluding remarks \label{summ}}

In the present paper, we studied four-dimensional topological Born-Infeld (BI)
charged black hole solutions in the context of Einsteinian cubic gravity
(ECG) in the presence of a bare cosmological constant. ECG is the most
general gravity up to cubic order in curvature which is independent of
dimension and its linearized spectrum coincides with general relativity one.
Also, both theoretical (open superstring theory) and experimental
(photon-photon interaction experiments) evidences suggest nonlinear theories
of electrodynamics such as BI model.
On the other hand, topological black hole solutions are known to be dual to some thermal states in the holographic settings, with quite different thermodynamical features depending on the specific topology.
To the best of our knowledge, this is the
first study of topological black hole solutions in ECG with nonlinear electrodynamics.

We first introduced the action of the theory and obtained its corresponding field equations.
Integrating these field equations, we obtained the electromagnetic potential as
well as a second order differential equation for the metric function. Then, we
presented the total mass $M$ formula by employing Abbott-Deser-Tekin method
and showed that the total mass depends on cubic coupling in general. We obtain the
total charge $Q$ via Gauss formula as well. Moreover, we discussed the
conditions under which the model is unitary and perturbatively ghost-free and explored
different possible cases. We found that if the bare cosmological constant
vanishes, then the model is unitary only if the solution is asymptotically flat.
Therefore, we focused on studying the thermodynamics of asymptotically flat solutions. These are necessarily of spherical topology. Expanding the metric function at the near horizon region up to second order in the field equations, we obtained the temperature $T$ and the total mass as functions
of horizon radius $r_{h}$ and charge $q$.
Next, we revealed the influences of cubic coupling $\lambda $, nonlinearity parameter $b$ and
black hole charge on the behavior of the temperature. We found that the
temperature decreases as the values of these parameters increase. It is
remarkable to mention that, as $b$ grows, the linear Maxwell
electrodynamics is reproduced. In addition, the temperature of spherical
asymptotically flat solutions have a maximum value as horizon radius
changes. This maximum temperature is smaller for greater values of $\lambda $, $%
b $ and $q$. In order to check the satisfaction of thermodynamics first law,
we turned to compute the entropy and the electric potential. We used the Wald
formula in order to obtain the entropy of our solutions. We showed that the
first law of thermodynamics is satisfied for spherical BI charged
asymptotically flat solutions of ECG.

Finally, we analysed thermal stability
of our solutions in both canonical and grand canonical ensembles. We showed
that the values of $q$ and $r_{h}$ specify how the stability is affected by
other parameters, in both ensembles. For one type of solutions, the system
is unstable for $b$ values larger than a maximum value $b_{\max }$
(including linear Maxwell case) in canonical ensemble. In the small $b$ regime
(highly nonlinear electrodynamics), there is also a minimum $\lambda $ value.
The system is stable if $\lambda \geq \lambda _{\min }$ for $b$
in the range $0$ to $b_{\max }$. There exists another type of solution with $\lambda _{\min }$
defined as before, but in which, black solutions are
stable if $\lambda \geq \lambda _{\min }$ with no upper bound for $b$.
Furthermore, for $\lambda $ values a bit smaller than $\lambda _{\min }$, the
system is stable for $b>b_{\min }$. There is also a critical value for $%
\lambda $ that if $\lambda <\lambda _{c}$, the black hole solutions are thermally
unstable for all $b$ values. In grand canonical ensemble, the first type
system is stable for large $b$ values (including linear Maxwell case). Nevertheless, in small $b$ regime, it is unstable for some $\lambda$ values. In this regime, the system becomes stable for $b$ values between $0$ and an upper bound, as $\lambda$ grows. The second type system, in grand canonical ensemble, is totally unstable for some $\lambda$ values. However, it becomes totally stable, as $\lambda$ increases.

In the present work, we focused on asymptotically flat solutions with vanishing
bare cosmological constant according to discussions given in Sec. \ref%
{consqu}. In such a scenario, the higher curvature terms have no contribution in the
total mass formula. However, if one considers possible (A)dS vacuum
solutions in a unitary model, a correction due to cubic curvature terms
appears in the total mass formula (see Eq. (\ref{hicumass})). This fact may
affect the thermdoynamical studies, drastically. It is fascinating to
explore these kinds of charged solutions in future works. In addition, here,
we considered Born-Infeld model as nonlinear electrodynamics. It is
interesting to study the effects of other known nonlinear models of electrodynamics on ECG
solutions. Moreover, it is worthwhile to explore the critical behavior of
nonlinearly charged solutions of ECG in both extended \cite{ext} and
non-extended \cite{nonext} thermodynamics phase spaces. Some of these issues
are under investigation and the results will appear elsewhere.

\begin{acknowledgments}
We thank the referee for constructive comments which helped us improve the paper significantly.
We also would like to thank Shahid Chamran University of Ahvaz, Iran for
supporting this work. We gratefully acknowledge Bayram Tekin, Tahsin Cagri
Sisman, Yen Chin Ong and Ahmad Sheykhi for valuable comments on early
version of the manuscript. AK thanks Ghadir Jafari for continuous
encouragement and support during this work.
\end{acknowledgments}

\appendix

\section{Calculation of $\mathcal{P}$ \label{ap1}}

Here, we present the detailed calculations for obtaining $\mathcal{P}$ in
Eq. (\ref{P1}). From Eq. (\ref{P}), we have%
\begin{align}
\mathcal{P}=& 12R_{a\ b}^{\ c\ d}R_{c\ d}^{\ e\ f}R_{e\ f}^{\ a\
b}+R_{ab}^{cd}R_{cd}^{ef}R_{ef}^{ab}  \notag \\
& -12R_{abcd}R^{ac}R^{bd}+8R_{a}^{b}R_{b}^{c}R_{c}^{a}\,.
\end{align}%
The above equation, can be re-written as 
\begin{align}
\mathcal{P}=& 12R^{acbd}R_{cedf}R_{a\text{ }b}^{\text{ }e\text{ }%
f}+R_{ab}^{cd}R_{cd}^{ef}R_{ef}^{ab}  \notag \\
& -12R_{cd}^{ab}R_{a}^{c}R_{b}^{d}+8R_{a}^{b}R_{b}^{c}R_{c}^{a}\,.
\label{P11}
\end{align}%
Then, using 
\begin{equation*}
R^{acbd}R_{cedf}R_{a\text{ }b}^{\text{ }e\text{ }f}=R^{acbd}R_{a\text{ }b}^{%
\text{ }e\text{ }f}R_{cfde}+\frac{1}{4}R^{acbd}R_{ac}^{\text{ \ \ \ }%
ef}R_{bdef},
\end{equation*}%
Eq. (\ref{P11}) can be expressed as
\begin{align}
\mathcal{P}=& 12R^{acbd}R_{a\text{ }b}^{\text{ }e\text{ }%
f}R_{cfde}+3R^{acbd}R_{ac}^{\text{ \ \ \ }%
ef}R_{bdef}+R_{ab}^{cd}R_{cd}^{ef}R_{ef}^{ab}  \notag \\
& -12R_{cd}^{ab}R_{a}^{c}R_{b}^{d}+8R_{a}^{b}R_{b}^{c}R_{c}^{a}\,.
\end{align}%
Rearranging the above equation yields%
\begin{align}
\mathcal{P}=&
12R_{cd}^{ab}R_{af}^{ce}R_{be}^{df}+4R_{cd}^{ab}R_{ab}^{ef}R_{ef}^{cd} 
\notag \\
& -12R_{cd}^{ab}R_{a}^{c}R_{b}^{d}+8R_{b}^{a}R_{a}^{c}R_{c}^{b}\,,
\end{align}%
which is what we expressed in Eq. (\ref{P1}).

\end{document}